\documentclass[aps,prl,nofootinbib,showpacs,twocolumn,groupedaddress]{revtex4}
\usepackage{amsmath,bm}
\newcommand{\beq}{\begin{eqnarray}}
\newcommand{\eeq}{\end{eqnarray}}

\newcommand{\be}{\bm{\eta}}

\newcommand{\bp}{\bm{\pi}}
\newcommand{\J}{\bm{J}}

\newcommand{\p}{\bm{p}}

\newcommand{\hp}{\bm{\widehat{\p}}}

\newcommand{\x}{\bm{x}}

\newcommand{\0}{\bm{0}}

\newcommand{\bv}{\bm{\varphi}}

\newcommand{\bx}{\bm{\xi}}

\newcommand{\bs}{\bm{\sigma}}

\newcommand{\bc}{\bm{\chi}}

\newcommand{\hbv}{\bm{\widehat\varphi}}

\newcommand{\hbxi}{\bm{\widehat\xi}}

\newcommand{\ar}{\stackrel{\hspace{0.04truecm}grav. }{\mbox{$\longrightarrow$}}}

\begin{document}

\title{Fermions, bosons, and locality in special relativity with two 
invariant scales }

\author{D. V. Ahluwalia-Khalilova}\email{d.v.ahluwalia-khalilova@heritage.reduaz.mx}
\homepage{http://heritage.reduaz.mx}

\affiliation{Theoretical Physics Group,
Facultad de Fisica, \\ Univ. Aut. de Zacatecas, Ap. Postal C-600,
Zacatecas 98062, Mexico\footnote{Present address: Department of Mathematics,
University of Zacateacs, Zacatecas, ZAC 98060, Mexico}}

\date{\today}

\begin{abstract}
We present a Master equation for description of fermions and bosons
for special relativities with two invariant 
scales ($c$ and $\lambda_P$). 
We introduce canonically-conjugate  
variables $\left(\chi^0, \bc\right)$ to 
$(\epsilon,\pi)$ of Judes-Visser. 
Together, they bring in a formal element of linearity and 
locality in an otherwise non-linear and non-local theory.
Special relativities with two invariant scales provide \textit{all} 
corrections, say, to the standard model of the high energy physics, 
in terms of \textit{one} 
fundamental constant, $\lambda_P$. It is emphasized that 
spacetime of special relativities with two invariant scales 
carries an intrinsic quantum-gravitational character. 

\end{abstract}


\maketitle


\textit{Introduction.}\textemdash~
There is a growing theoretical evidence that gravitational
and quantum frameworks carry some elements of incompatibilities. 
The question is how deep are the indicated changes, and 
what precise form they may take. One hint comes from 
the observation that incorporating gravitational effects in 
quantum measurement of spacetime events leads to a Planck-scale
saturation. In the framework of Kempf, Mangano,  Mann, and one of us
\cite{ak,dva_pla}, 
the gravitationally-induced modification to the 
de Broglie (dB) wave-particle duality 
takes the form \cite{dva_pla} 
\beq
\quad \lambda_{dB}=\frac{h}{p}
\quad \ar\quad
\lambda = \frac{\overline{\lambda}_P}
{\tan^{-1}(\overline{\lambda}_P/\lambda_{dB})},
\eeq
where $\overline{\lambda}_P$ is the Planck circumference ($=2\pi\lambda_P$), 
with $
\lambda_P = \sqrt{\hbar G /c^3}$.
The $\lambda$ reduces to $\lambda_{dB}$ for
the low energy regime, and saturates to $4 \lambda_P$ in the 
Planck realm. In this way the Planck scale is not merely 
a dimensional parameter but has been brought in relation
to a universal saturation of
gravitationally-modified de Broglie wavelengths. 

This is a very welcome situation for theories of
quantum gravity where for a long time a paradoxical
situation had existed \cite{dsr2,struc2}. 
Each inertial observer
could measure in his frame the fundamental universal 
constants, $\hbar,c,G$, and obtain from  
them a universal  fundamental constant, $\lambda_P$. 
And yet this very $\lambda_P$ \textendash~ being a length scale
\textendash~
is subject to special-relativistic length contraction
which  paradoxically makes it loose its universal character.

The indicated saturation then not only resolves this paradoxical
situation but also suggests that special relativity must suffer
a modification. This modification
must be endowed with the property that it carries two invariant
scales; one the usual $c$, and the second $\lambda_P$.

Amelino-Camelia, followed by Magueijo and Smolin, and Judes and Visser 
\cite{dsr1,dsr2,jv}, have provided
first steps towards development of a special relativity with 
two invariant scales (\texttt{SR2}); 
while Lukierski, Nowicki, and Kowalski-Glikman
\cite{struc1,struc2} have brought to attention the underlying 
quantum/Hopf-group structure \cite{qg} of such theories.\footnote{Instead 
of the term ``doubly special relativity''
coined in the work of Amelino-Camelia\cite{dsr1},
we prefer to use the phrase ``special relativity with two 
invariant scales.'' Without in any way 
questioning  physics content of Amelino-Camelia's proposal,
we take this non-semantic issue  
for the following reason.   The special 
of ``special relativity'' refers to the circumstance that 
one restricts to a special class of inertial observers which move
with a relative uniform velocity. The general of ``general relativity''  
lifts this restrictions. The ``special'' of special relativity
has nothing to do with one versus two
invariants scales. It rather refers to the special class of inertial 
observers; a circumstance that remains unchanged in  special relativity
with two invariant scales. The theory of general relativity with
two invariant scales would thus not be called ``doubly general relativity.''}
The necessity for a  SR2 as argued in Refs.  \cite{dsr2,struc2} is
similar to ours, while motivation of Ref.  \cite{dsr1} is contained in
certain anomalies in astrophysical data \cite{d1,d2,d3,d4,d5}. 
Simplest of SR2 theories result from
keeping the algebra of boost- and 
rotation- generators
intact while modifying the boost parameter in a non-linear manner. 
Specifically, in the \texttt{SR2} of Amelino-Camelia
the boost parameter, $\bv$, changes from the special relativistic form
\beq
\cosh{\varphi} = \frac{E}{m }\,,\quad
\sinh{\varphi} = \frac{p}{m}\,,\quad
\hbv=\frac{\p}{p}\,. \label{dirac}
\eeq
to a new structure \cite{gak,jv}
\begin{subequations}
\beq
\cosh\xi &=& 
\frac{1}{\mu}\left( \frac{
e^{\lambda_P  E} 
-\cosh\left(\lambda_P\, m \right)}
{\lambda_P \cosh\left( \lambda_P \,m/2\right)}
\right)\,,\label{gac1}\\
\sinh\xi &=& 
\frac{1}{\mu}\left(
\frac{p \,
e^{\lambda_P  E}
}
{\cosh\left(\lambda_P \,m/2\right)} \right) \,,
\quad
\hbxi=\frac{\p}{p}\,, \label{gac2}
\eeq
\end{subequations}
while for the \texttt{SR2}  of  Magueijo and Smolin the change takes
the form \cite{jv,dsr2}
\begin{subequations}
\beq
\cosh\xi &=& 
\frac{1}{\mu}
\left(\frac{E}{1-\lambda_P\,E}\right)
\,,\label{ms1}\\
\sinh\xi &=& 
\frac{1}{\mu}
\left(\frac{p}{1-\lambda_P\,E}\right)\,,
\quad
\hbxi=\frac{\p}{p}
\,. \label{ms2}
\eeq
\end{subequations}
Here,  $\mu$ is a Casimir invariant of \texttt{SR2} (see Eq. 
(\ref{ci}) below) and is given by
\beq
\mu =
\begin{cases}
\frac{2}{\lambda_P}\,
\sinh\left(
\frac{\lambda_P \,m}{2}
\right) & \mbox{for  Ref. \cite{dsr1}'s \texttt{SR2} }\\
\frac{m}{1-\lambda_P m}
&       \mbox{for  Ref. \cite{dsr2}'s \texttt{SR2} }
\end{cases}
\eeq
The notation is that of Ref. \cite{jv}; with the minor  exceptions:
$\lambda$, $\mu_0$, $m_0$ there are $\lambda_P$,  $\mu$, $m$ here. 
In what follows we shall 
\textit{generically} represent boost parameter associated
with special relativities with one, or two, invariant scales by $\bx$.
The former relativity shall be abbreviated as \texttt{SR1} 
(to distinguish it from
\texttt{SR2}).\footnote{In this notation the Galilean relativity is 
denoted by \texttt{SR0}.}
Note that giving the explicit expressions for both the 
$\sinh\xi$ and $\cosh\xi$ in Eqs.~(\ref{gac1},\ref{gac2})
is necessary in order to fix the form of the energy-momentum dispersion 
relation through the identity: $\cosh^2\xi-\sinh^2\xi=1$. 
Of course, one may have chosen to work
in terms of one of the hyperbolic trigonometric functions 
\textit{and} the dispersion relation, instead.

At this early stage it is not clear if
there is a unique \texttt{SR2}, or, if 
the final choice will be eventually settled by observational data,
or by some yet-unknown physical principle. Given this ambiguity, this
\textit{Letter} addresses itself to presenting a Master equations 
for fermionic and bosonic representations for generic \texttt{SR2}.

\textit{Master equation for spin-1/2: Dirac case.}\textemdash~ 
Since the underlying 
spacetime symmetry generators remain unchanged much of the  
formal apparatus of the finite dimensional representation
spaces associated with the Lorentz group remains intact.
In particular, there still exist $(1/2,\,0)$ and $(0,\,1/2)$
spinors. But now they transform from the rest frame, to
an inertial frame in which the particle has momentum, $\p$ as:
\begin{subequations}
\beq
\phi_{(1/2,\,0)}\left(\p\right)
& = & \exp\left( + \frac{\bs}{2}\cdot\bx \right)
\phi_{(1/2,0)}\left(\0\right)\,\label{a}\\
\phi_{(0,\,1/2)}\left(\p\right)
& = & \exp\left(- \frac{\bs}{2}\cdot\bx \right)
\phi_{(1/2,0)}\left(\0\right)\,.\label{b}
\eeq  
\end{subequations}
Since in this \textit{Letter}
we do not undertake a study of the behavior of these spinors under
the parity operation, or examine the  massless limit in detail, 
we do not identify the 
$(0,\,1/2)$ spinors as \textit{left-handed} and the $(1/2,\,0)$ spinors
as \textit{right-handed.} Since 
the null momentum vector $\0$ is still isotropic,
one may assume that (see p. 44  of 
Ref. \cite{Ryder} \textit{and} Refs. \cite{dva_review,gg}): 
\beq
\phi_{(0,1/2)}\left(\0\right) = \zeta \,\phi_{(1/2,0)}\left(\0\right)\,,
\label{c}
\eeq
where $\zeta$ is an undetermined phase factor. 
The analysis presented in Ref. \cite{ka} also  convinces
us that the validity of the identity (\ref{c}) is independent of the 
``right-left'' identification of the standard argument 
\cite{Ryder,dva_review,gg}.
In general, the phase
$\zeta$ encodes
C, P, and T properties. The interplay of
Eqs. (\ref{a}-\ref{b}) and (\ref{c}) yields the Master equation for
the $(1/2,\,0)\oplus(0,\,1/2)$ spinors,
\beq
\psi\left(\p\right) = \left(
			\begin{array}{c}
			\phi_{(1/2,\,0)}\left(\p\right)\\
			\phi_{(0,\,1/2)}\left(\p\right)
			\end{array}
		     \right)\,,
\eeq
to be
\beq
\left(
\begin{array}{cc}
-\zeta & \exp\left(\bs\cdot\bx\right) \\
\exp\left(- \bs\cdot\bx\right)\psi\left(\p\right) & - \zeta^{-1}
\end{array}
\right) \psi\left(\p\right) = 0\,.\label{meq}
\eeq
This is one of the central results of this \textit{Letter}.

As a check, taking $\bx$ to be $\bv$, and after some
simple algebraic manipulations,
the Master equation (\ref{meq}) reduces to:
\beq
\left(
 	\begin{array}{cc}
	- m \zeta & E \openone_2 + \bs\cdot \p \\
	E \openone_2 - \bs\cdot \p & - m \zeta^{-1}
	\end{array} 
\right) \,
\psi\left(\p\right) = 0\,,\label{d}
\eeq
where $\openone_n$ stands for $n\times n$ identity matrix
(and $0_n$ shall represent the corresponding null matrix) .
With the given identification of the boost parameter
we are in the realm
of \texttt{SR1}. There, the operation of
parity is well understood. Demanding  parity
covariance for Eq. (\ref{d}), we obtain
$\zeta=\pm 1$. Identifying 
\beq
\left(  \begin{array}{cc}
	0_2 & \openone_2 \\
	\openone_2 & 0_2
	\end{array}
\right)\,,\quad
\left(  \begin{array}{cc}
	0_2 & -\bs\\
	\bs & 0_2
	\end{array}
\right)\,,
\eeq
with the Weyl-representation $\gamma^0$, and $\gamma^i$, respectively,
Eq. (\ref{d}) reduces to the Dirac equation of  \texttt{SR1} 
\beq
\left(\gamma^\mu p_\mu \mp m\right)\psi\left(\p\right)=0\,.\label{de}
\eeq
The linearity of the Dirac equation in, $p_\mu= (E,-\p)$, is now clearly
seen to be associated with two observations: 

\begin{enumerate}
\item[$\mathcal{O}_1$.]
That, $\bs^2 = \openone_2$; and
\item[$\mathcal{O}_2$.]
That in \texttt{SR1}, the hyperbolic functions
\textendash~ 
see Eq. (\ref{dirac}) \textendash~ associated with the boost parameter
are linear in $p_\mu$. 
\end{enumerate}
In \texttt{SR2}, observation $\mathcal{O}_1$
still holds. But, as Eqs. (\ref{gac1} - \ref{gac2}) 
show, $\mathcal{O}_2$ is strongly violated. For this reason the Master equation
(\ref{meq}) cannot be cast in a manifestly covariant form with
a finite number of contracted Lorentz indices of \texttt{SR2} 
as long as we mark spacetime events by $x^\mu$ of \texttt{SR1}.

The last inference is also a welcome result as it indicates
a possible intrinsic non-locality in \texttt{SR2}s. Since in all 
\texttt{SR2}s
the shortest spatial length scales that can be probed are bound
from below by $\lambda_P$, the naively-expected $\delta^3\left(\x-
\x^\prime\right)$ in the anticommutators of the form 
$\left\{\Psi_i\left(\x,t\right),\,\Psi^\dagger_j\left(\x^\prime, t
\right)\right\}$ should be replaced by an highly, but
not infinitely, 
peaked Gaussian-like functions with half-width of the order of
$\lambda_P$.

The extension of the presented formalism for Majorana spinors 
is more subtle \cite{m1,m2,m3}. 
We hope to present it an extended version of this 
\textit{Letter.}

\textit{Master equation for higher spins.}\textemdash~ 
The above-outlined procedure applies
to all, bosonic as well as fermionic,  
$(j,0)\oplus(0,j)$ representation spaces. It is 
not confined to $j=1/2$. A straightforward generalization 
of the $j=1/2$ analysis immediately yields the Master equation 
for an arbitrary-spin,
\beq
\left(
\begin{array}{cc}
-\zeta & \exp\left(2\J\cdot\bx\right) \\
\exp\left(- 2\J\cdot\bx\right)\psi\left(\p\right) & - \zeta^{-1}
\end{array}
\right) \psi\left(\p\right) = 0\,,\label{j}
\eeq
where
\beq
\psi\left(\p\right) = \left(
			\begin{array}{c}
			\phi_{(j,\,0)}\left(\p\right)\\
			\phi_{(0,\,j)}\left(\p\right)
			\end{array}
		     \right)\,.\label{js}
\eeq
Equation (\ref{j}) 
contains the central result of the previous section as a 
special case.  
For studying the \texttt{SR1} limit it is convenient to bifurcate the
$(j,0)\oplus(0,j)$ space into two sectors by splitting the 
$2(2j+1)$ phases, $\zeta$,
into two sets: $(2j+1)$ phases $\zeta_+$, and the other  
$(2j+1)$ phases $\zeta_-$. Then,  in particle's rest frame
the $\psi(\p)$ may be written as:
\beq
\psi_h(\0)=
\begin{cases}
u_h(\0) &  \mbox{when}~ \zeta=\zeta_+\cr
v_h(\0) &  \mbox{when} ~\zeta=\zeta_- 
\end{cases}
\eeq
The explicit forms of $u_h(\0)$ and $u_h(\0)$ which we 
shall use (see Eq. (\ref{c})) are:
\beq
u_h(0)=\left(
\begin{array}{c}
\phi_h(\0) \\
\zeta_+ \,\phi_h(\0)   
\end{array}
\right),\,
v_h(0)=
\left(
\begin{array}{c}
\phi_h(\0) \\
\zeta_-\, \phi_h(\0)   
\end{array}
\right),
\eeq
where the   $\phi_h(\0)$ are defined as:
$\J\cdot \hp\, \phi_h(\0) = h \,\phi_h(\0)$, and $h=-j,-j+1,\ldots,+j$.
In the parity covariant \texttt{SR1} 
limit, we find $\zeta_+ = +1$ while   
$\zeta_- = -1$.

As a check, for $j=1$, identification of $\bx$ with $\bv$,
and after implementing parity covariance, yields
\beq
\left(\gamma^{\mu\nu}p_\mu p_\nu \mp m^2\right)\psi(\p)=0\,.\label{bwweq}
\eeq 
The  $\gamma^{\mu\nu}$ are unitarily equivalent  
to those of Ref. \cite{bww}, and thus we reproduce 
\textit{bosonic matter fields}
with $\left\{C,\,P\right\} = 0$.  A carefully taken massless limit then shows
that the resulting equation is consistent with the free Maxwell equations
of electrodynamics.

Since the $j=1/2$ and $j=1$ representation spaces of \texttt{SR2} reduce to
the Dirac and Maxwell descriptions, it is apparent, that the \texttt{SR2}
contains physics beyond the linear-group realizations of \texttt{SR1}.
To the lowest order in $\lambda_P$,  Eq. (\ref{meq}) yields
\begin{subequations}
\beq
\left(
\gamma^\mu {p}_\mu + \tilde{m} + 
\delta_1\, \lambda_P\right)
\psi(\p)=0\,,
\eeq
where
\beq
\tilde{m}
&=& \left(
\begin{array}{cc}
-\zeta  & 0_2  \\
0_2 & -\zeta^{-1}
\end{array}
\right)\,m \,
\eeq
and
\beq
\delta_1 =
\begin{cases}
\gamma^0\left(\frac{E^2-m^2}{2}\right)+\gamma^i p_i\, E
& \mbox{for  Ref. \cite{dsr1}'s \texttt{SR2}}\\
\gamma^\mu p_\mu\, \left(E-m\right)
&       \mbox{for Ref. \cite{dsr2}'s \texttt{SR2}}
\end{cases}
\eeq
\end{subequations}
Similarly, the presented Master equation can be used  
to obtain \texttt{SR2}'s counterparts for Maxwell's electrodynamic.
Unlike the Coleman-Glashow framework \cite{cg}, the principle
of special relativity with two invariant scales provides \textit{all} 
corrections, say, to the standard model of the high energy physics, 
in terms of \textit{one} \textendash~ and \textit{not forty six} \textendash~ 
fundamental constant, $\lambda_P$.

\textit{Spin-1/2 and Spin-1 description in Judes-Visser 
Variables.}\textemdash~
We now take the tentative position,
that the ordinary energy-momentum $p^\mu$ is not the natural physical
variable in \texttt{SR2}s. The Judes-Visser 
variables \cite{jv}: $\eta^\mu \equiv 
\left(\epsilon(E,p),\,\bp(E,p)\right)=(\eta^0,\be)$
appear more suited to describe physics sensitive to Planck scale.
The $\epsilon(E,p)$ and $\bp(E,p)$ 
relate to the rapidity parameter $\bx$ of  \texttt{SR2} 
in same functional form
as do $E$ and $\p$ to  $\bv$ of \texttt{SR1}:
\beq
\cosh\left(\xi\right)= \frac{\epsilon(E,p)}{\mu}\,,\quad
\sinh\left(\xi\right)= \frac{\pi(E,p)}{\mu}\,,
\eeq
where 
\beq
\mu^2= \left[\epsilon(E,p)\right]^2 - \left[\bp(E,p)\right]^2\,.
\label{ci}
\eeq
They provide 
the most economical and physically transparent formalism for 
representation space theory in \texttt{SR2}.   
For $j=1/2$ and $j=1$, Eq. (\ref{j}) yields the \textit{exact}
SR2 equations for $\psi(\bp)$:
\beq
&&\left(\gamma^\mu \eta_\mu + \tilde{\mu}\right)
\psi\left(\bp\right)=0\,,\label{denew}\\
&&\left(\gamma^{\mu\nu}\eta_\mu \eta_\nu + \tilde{\mu}^2\right)
\psi(\bp)=0\,,\label{bwweqnew}
\eeq
where
\beq
\tilde{\mu}
&=& \left(
\begin{array}{cc}
-\zeta^{-1}  & 0_2 \\
0_2 & -\zeta
\end{array}
\right)\,\mu\,.
\eeq

\textit{Concluding Remarks.}\textemdash~
Our task in this \textit{letter} was to provide a description 
of fermions and bosons at the  level of representation space 
theory in \texttt{SR2}. However, we confined entirely to the
representations of the type $(j,0)\oplus(0,j)$ \textendash~ 
these types are important for matter fields, and 
to study gauge-field strength tensors.
To study \texttt{SR2}'s effect on the gauge fields and  weak-field gravity
the present \textit{Letter's} formalism needs to be extended to 
$(j,j)$ representation spaces.
In view of Weinberg's earlier works \cite{sw1964} it is known
that there is a deep connection between \textit{local} quantum field theory,  
\textit{SR1}  $(j,j)$  spaces \cite{ka}, and the \textit{equality} 
of the inertial and gravitational masses. Therefore, the suggested study
must answer \texttt{SR2}'s effect on the equivalence principle.

In quantum field theoretic framework, 
the  special relativity's spacetime $x^\mu$ is canonically conjugate
to $p_\mu$, and appears in the  field operators as:
\beq
\Psi(x)= \int \frac{d^3 \p}{(2\pi)^3} \frac{m}{p_0} 
\sum_{h=-j}^{+j}{\Big[}
a_h(\p) u_h(\p) e^{- i p_\mu x^\mu} &&\nonumber \\ +\;
b_h(\p) v_h(\p)
e^{i p_\mu x^\mu}{\Big]}\,,&&
\eeq
where the particle-antiparticle spinors, $u_h(\p)$
and $v_h(\p)$ (generically represented by $\psi_h(\p)$), 
are solutions of the Master equations (but with 
$\bx\rightarrow\bv$)
introduced above, and can be readily obtained from:
\beq
\psi_h(\p)=
\left(
\begin{array}{cc}
\exp(+ \J\cdot\bv) & 0_{2j+1} \\
 0_{2j+1} & \exp(-\J\cdot\bv)
\end{array}
\right)\psi_h(\0)\,.
\eeq
Now, as our discussion on non-locality indicates 
$x^\mu$ of  \texttt{SR1} is perhaps not the natural 
physical spacetime variable
at the Planck scale. The spacetime at Planck scale, we suggest,
is represented by new event vectors $\chi^\mu$ (to be treated as
``canonically conjugate'' to Judes-Visser variable $\eta_\mu$); and
suggests the following definition for the field operators
built upon the SR2's spinors:
\beq
\Psi(\chi)= \int \frac{d^3 \be}{(2\pi)^3} \frac{\mu}{\eta_0} 
\sum_{h=-j}^{+j}{\Big[}
a_h(\be) u_h(\be)
e^{- i \eta_\mu \chi^\mu}  &&\nonumber \\ +\;
b_h(\be) v_h(\be)
e^{i \eta_\mu \chi^\mu}{\Big]}\,,&&
\eeq
with 
\beq
\psi_h(\be)=
\left(
\begin{array}{cc}
\exp(+ \J\cdot\bx) & 0_{2j+1} \\
 0_{2j+1} & \exp(-\J\cdot\bx)
\end{array}
\right)\psi_h(\0)\,.
\eeq
Immediately, we verify that for spin-$1/2$ fermions in SR2
\beq
\left\{\Psi_i\left(\bc,\chi^0\right),\,\Psi^\dagger_j\left(\bc^\prime,
\chi^0\right)\right\} 
=\delta^3
\left(\bc-\bc^\prime\right)\,\delta_{ij}\,.
\eeq
What appears as non-locality in the space of events marked 
by $x^\mu$ now,  in the space of events marked by 
$\chi^\mu$,  exhibits itself as
locality.  This is a rather unexpected observation and it
calls for a deeper understanding of 
the $\eta_\mu$ and $\chi^\mu$ description of SR2. 
The Planck length is intrinsically built in the
latter spacetime variables, and it may carry significant relevance for
extending SR2 to the gravitational realm. 

The evolution of special relativity in the sequence\footnote{The symbols 
above the arrows indicate the invariants for the subsequent 
\texttt{SRn}.}  
\beq
\mbox{\texttt{SR0}}
\stackrel{c}{\longrightarrow}
\mbox{\texttt{SR1}}
\stackrel{c,\lambda_P}{\longrightarrow}
\mbox{\texttt{SR2}}
\eeq
 translates to giving spacetime, first, a 
\textit{relativistic} and, then, a \textit{quantum-gravitational} 
character. The work initiated here, and in Ref. \cite{aa}, gives concrete
shape to modifications that one may expect in the standard model
of high-energy physics and theory of gravitation.

\begin{acknowledgments}
\textit{Acknowledgments.}\textemdash~
One of us (DVA) thanks Giovanni Amelino-Camelia, 
Gaetano Lambiase, and their respective institutes
in Rome and Salerno, for discussions and hospitality
in May 2001. It is also our pleasure to record that part
of this work was initiated during those visits, and 
that concurrently  Arzano and Amelino-Camelia  \cite{aa}
have obtained some results similar to the ones 
reported here. We are also grateful to Giovanni for freely
exchanging with us his thoughts on \textit{Doubly Special Relativity}
and for engaging in  time-consuming  
extended discussions on the material presented here.

This work is supported by Consejo Nacional de Ciencia y
Tecnolog\'ia (CONACyT, Mexico).
\end{acknowledgments}

\end{document}